\documentclass[aps,prc,superscriptaddress,10pt,showpacs,notitlepage,  
nofootinbib
]{revtex4-1}
\usepackage{bm}
\usepackage{amsmath,amssymb,amsthm}
\usepackage{latexsym,graphicx,color}
\usepackage{enumerate}
\usepackage{amssymb}
\usepackage{hyperref}
\usepackage{mathtools}

\usepackage{graphicx}
\graphicspath{ {images/} }

\usepackage{color}

\def\lesssim{\ \raise.3ex\hbox{$<$}\kern-0.8em\lower.7ex\hbox{$\sim$}\ }
\def\gesim{\ \raise.3ex\hbox{$>$}\kern-0.8em\lower.7ex\hbox{$\sim$}\ }

\begin{document}
\title{Strong-coupling effects of pairing fluctuations, and Anderson-Bogoliubov mode in neutron $^1S_0$ superfluids in neutron stars}
\author{Daisuke Inotani}
\email{dinotani@keio.jp}
\affiliation{Department of Physics $\&$ Research and Education Center for Natural Sciences,\\ Keio University,Hiyoshi 4-1-1, Yokohama, Kanagawa 223-8521, Japan}
\author{Shigehiro Yasui}
\email{yasuis@keio.jp}
\affiliation{Department of Physics $\&$ Research and Education Center for Natural Sciences,\\ Keio University,Hiyoshi 4-1-1, Yokohama, Kanagawa 223-8521, Japan}
\author{Muneto Nitta}
\email{nitta(at)phys-h.keio.ac.jp}
\affiliation{Department of Physics $\&$ Research and Education Center for Natural Sciences,\\ Keio University,Hiyoshi 4-1-1, Yokohama, Kanagawa 223-8521, Japan}
\date{\today}
\begin{abstract}
We investigate effects of thermal and quantum fluctuations of the superfluid order parameter in $^1S_{0}$ superfluids in neutron stars.
We construct a separable potential to reproduce the $^1S_{0}$ phase shift reconstructed  by using the partial wave analysis from nucleon scattering data.
We include superfluid fluctuations within a strong-coupling approximation developed by Nozi\`eres and Schmitt-Rink and determine self-consistently the superfluid order parameter as well as the chemical potential. We show that the quantum depletion, which gives a fraction of noncondensed neutrons at zero temperature due to quantum pairing fluctuations, plays an important role not only near the critical temperature from superfluid states to normal states but also at zero temperature. We derive the dispersion relation of Anderson-Bogoliubov modes associated with phase fluctuations, and show also that there is a nonzero fraction of noncondensed components in the neutron number as a result of the strong-coupling effect.
Our results indicate that superfluid fluctuations are important for thermodynamic properties in neutron stars. 
\end{abstract}
\maketitle

\section{Introduction}
Neutron stars are important astrophysical objects for studies of properties of nuclear matter at high density, with rapid rotation, strong magnetic field and so on, whose environments are quite different from those in normal nuclei (see Refs.~\cite{Graber:2016imq,Baym:2017whm} for recent reviews).
It was recently reported that there are massive neutron stars whose masses are almost twice as large as the solar mass~\cite{Demorest:2010bx,Antoniadis1233232} and it was also observed that the gravitational waves were emitted from a binary neutron star merger~\cite{TheLIGOScientific:2017qsa}.
Neutron stars are interesting also as macroscopic laboratories for studying  quantum effects in 
high density matter. 
Inside neutron stars, one of the most important ingredients are pairing phenomena induced by the attractive force between two nucleons lying near the Fermi surface in momentum space~\cite{Dean:2002zx}.
In the literature, there are many studies on the neutron superfluidity and proton superconductivity for explaining the observation of the neutron stars (see Refs.~\cite{Chamel2017,Haskell:2017lkl,Sedrakian:2018ydt} for recent reviews).
For example, it was expected that low-energy excitation modes in superfluidity and superconductivity are important to explain  pulsar glitches, i.e., sudden speed-up of rotations of neutron stars~\cite{Baym1969,Pines1972,Takatsuka:1988kx}.
Such excitation modes can affect an enhancement of neutrino emissivities from neutron stars~\cite{Yakovlev:2000jp,Potekhin:2015qsa,Yakovlev:1999sk,Heinke:2010cr,Shternin2011,Page:2010aw}.\footnote{We comment that pulsar glitches may also be explained by the existence of quantized vortices in superfluids~\cite{reichley,Anderson:1975zze}}

At present, it is considered that the vast region of uniform neutron matter exists at low density under the crust region near the surface of neutron stars~\cite{Chamel:2008ca}.
Since early studies, it has been theoretically proposed that neutron $^1S_0$ superfluid states are realized at low-density region, where the effective attraction between two neutrons is dominated by the  $^1S_0$ interaction (spin singlet, $S$ wave, and zero total spin)\cite{Migdal:1960} (see also Ref.~\cite{Dean:2002zx} and the references therein).
\footnote{The $^1S_0$ interaction turns to be repulsive due to the strong core repulsion at higher densities~\cite{1966ApJ...145..834W}. Instead, the dominant interactions are provided by the $^{3}P_{2}$ interaction at high density, leading to the $^{3}P_{2}$ superfluidity which is described by the Bogoliubov-de Gennes (BdG) equation (see Ref.~\cite{PhysRevResearch.2.013194} and references therein),
 and also by the Ginzburg-Landau (GL) equation as the low-energy effective theory of the BdG equation (see Ref.~\cite{PhysRevC.101.025204} and references therein).
}

So far, the neutron 
superfluidity has been discussed mostly in terms of the mean-field theory. 
Recently, however, the importance of effects of pairing fluctuations in neutron $^1S_0$ superfluid states has been theoretically pointed out in the context of the Bardeen-Cooper-Schrieffer (BCS)--Bose-Einstein-condensation (BEC) crossover phenomenon known in ultracold atomic Fermi gases~(see Refs.~\cite{Strinati:2018wdg,doi:10.1146/annurev-nucl-102014-021957,Stein1995,PhysRevC.88.054315,PhysRevC.80.045802,PhysRevC.81.025803,PhysRevC.82.024911}).
At low density, the $S$-wave interaction between two neutrons is well described by the effective range expansion (ERE) with the negative scattering length $a_s=-18.8\pm0.3$ ${\rm fm}$ and the effective range $r_{\rm {eff}}=2.75\pm0.11$ ${\rm fm}$~\cite{Dean:2002zx}.
Thus, for a typical Fermi momentum $k_{\rm F}\simeq 1$ ${\rm fm}^{-1}$ in neutron stars, the strength of the pairing interaction is given by a nondimensional parameter, $(k_{\rm F} a_s)^{-1}\simeq -0.05$.
The large magnitude of $k_{\rm F} a_s$ implies that, as long as effects of the finite effective range are negligibly small, properties of the $^1S_0$ superfluid are expected to be similar to ones in dilute two-component (pseudospin up and down) atomic Fermi gas in the crossover regime, where superfluid fluctuations become remarkably large.
Thus, neutron $^{1}S_{0}$ superfluids should be regarded as a strongly coupled system.
In the condensed matter physics, it is known that effects of pairing fluctuations beyond the mean-field approximation can be described by the Nozi\`eres and Schmitt-Rink (NSR) scheme~\cite{Nozieres1985}.
The NSR scheme is applicable semi-quantitatively 
to the BCS-BEC crossover phenomena in cold atom physics~\cite{RevModPhys.82.1225,Mueller_2017,Jensen2019,PhysRevLett.71.3202}.
Motivated by this success, the NSR scheme has been adopted to studies of strong-coupling properties of 
nuclear systems above the superfluid transition temperature $T_{\rm c}$~\cite{Strinati:2018wdg,PhysRevC.88.054315,PhysRevC.82.024911,Stein1995}.
Recently, effects of pairing fluctuations in the $^1S_0$ neutron superfluid phase have been studied for equation-of-state in neutron matter~\cite{PhysRevA.97.013601} by considering the finite effective range as well as a strong-coupling effects in the NSR scheme.
We notice, however, that the ERE is broken down at high momentum where the $^1S_0$-channel interaction becomes repulsive, and it turns out that the phase transition from superfluids to normal states cannot be described. 
Thus, it is necessary to make a more precise effective potential for further quantitative understanding of 
superfluidity in a wide range of density regions in neutron stars. 
This problem can be overcome by considering an effective separable potential with a cutoff function~\cite{Tajima:2019saw}.
The explicit form of the cutoff function is numerically determined to reproduce the phase shift in the $^1S_0$-channel from experimental data of nucleon scattering in a given momentum range.
In Ref.~\cite{Tajima:2019saw}, the separable potential form was applied to study pairing fluctuations in the NSR scheme. However, they discussed the effects of pairing fluctuations only in the normal phase above $T_{\rm c}$.

In this paper, we extend the NSR scheme to the neutron superfluid below  $T_{\rm c}$ described by a nuclear potential, which reproduces scattering phase shifts in the $^1S_0$-channel. With this setup, we investigate the gap strength in neutron $^{1}S_{0}$ superfluids covering a wider range of the density and temperature from zero to $T_{\rm c}$.
We determine simultaneously the gap strength as well as the chemical potential, where the chemical potential is much affected by the strong-coupling effect.
We find that our numerical results are quantitatively different from those obtained within the mean-field theory which is valid only in the weak-coupling limit.

One of the advantages of the NSR scheme in the superfluid phase is to explicitly treat the noncondensed neutron pairs consisting of the bosonic collective excitations, which is known as Anderson-Bogoliubov (phase, sound, or phonon) and Higgs (or amplitude) modes. From the spectral functions of these modes, we show that in the $^1S_0$ neutron superfluidity, the Anderson-Bogoliubov mode plays a remarkable role not only near the critical temperature but also at zero temperature. Those gapless-collective modes will affect the transport properties in the neutron stars, e.g., the cooling process by neutrino emissions (see Refs.~\cite{Graber:2016imq,Baym:2017whm} and references therein). Our result indicates that effects of the quantum fluctuations, which have been usually ignored in the most of previous theoretical works, are crucial to describe superfluid properties of neutron star interiors.

This paper is constructed as follows.
In Sec.~\ref{sec:formalism}, we describe briefly our formalism to construct the neutron-neutron potential and explain the gap equations for the neutron $^{1}S_{0}$ superfluid including the fluctuation effects in the NSR scheme.
In Sec.~\ref{sec:result}, we show our numerical results about the gap strength and chemical potential.
We show the dispersion relations of Anderson-Bogoliubov and Higgs modes as fluctuation modes, and also that there is a nonzero fraction of noncondensed components in the neutron number as a result of the strong-coupling effect.
Finally, Sec.~\ref{sec:summary} is devoted to our conclusion and outlooks.

\section{Formalism}\label{sec:formalism}
\begin{figure}
\centerline{\includegraphics[width=9cm]{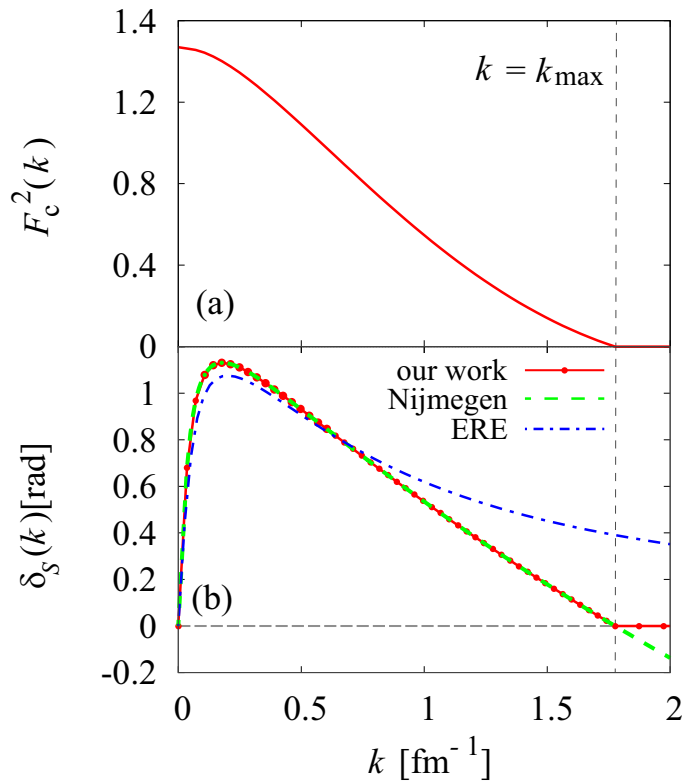}}
\caption{(Color online) (a) Calculated from factor $F_{\rm c}(k)$ and (b) reconstructed $^1S_0$-channel phase shift $\delta_S(k)$ (solid line and dots). The dotted and chain line show $\delta_S(k)$ estimated by using the partial wave analysis from the scattering experimental data~\cite{Dean:2002zx} and calculated within the ERE, respectively, for comparison.}
\label{Fig:ps}
\end{figure}
\par
We consider a neutron matter with a $^1S_0$ interaction, described by the Hamiltonian
\begin{align}
H&=
\sum_{{\bm k}\sigma} 
\xi_{k}
c^{\dagger}_{{\bm k}\sigma}c_{{\bm k}\sigma}
+\sum_{\bm {k},\bm {k}',\bm {q}}
V_S\left( \bm {k},\bm {k}' \right)
c^{\dagger}_{{\bm k}+\frac{\bm q}{2}\uparrow}
c^{\dagger}_{-{\bm k}+\frac{\bm q}{2}\downarrow}
c_{-{\bm k}'+\frac{\bm q}{2}\downarrow}
c_{{\bm k}'+\frac{\bm q}{2}\uparrow},
\label{Eq:Ham}
\end{align}
where $c_{{\bm k}\sigma}$($c^\dagger_{{\bm k}\sigma}$) is the annihilation (creation) operator of a neutron with the momentum ${\bm k}$ and the spin $\sigma=\uparrow, \downarrow$. $\xi_{k}=k^2/(2m)-\mu$ is the kinetic energy measured from the chemical potential $\mu$, where $m$ is the neutron mass. We assume that the interaction between neutrons is given by the $^1S_0$ pairing interaction, whose form is described by an attractive separable potential as
\begin{align}
V_S\left( {\bm k},{\bm k}' \right)=-U_S F_{\rm c}(k)F_{\rm c}(k'),
\end{align}
where $U_S > 0$ and $F_{\rm c}(k)$ are the coupling constant and the form factor, respectively. 
These are related to the $^1S_0$ phase shift $\delta_S(k)$ as 
\begin{align}
k\cot \delta_S \left(k \right)&=\frac{4\pi}{m}\left[ U_S^{-1}+{\rm {Re}}\Pi_S\left( k \right)\right]F^{-2}_{\rm c}\left( k \right),
\label{Eq:pseq}
\end{align} 
where
\begin{align}
\Pi_S\left( k \right)
&=
\sum_{\bm k'}
\frac{F_c^2\left( k' \right)}
{\frac{k^2}{m}+\frac{k'^2}{m}+i0},
\end{align}
is the two-body correlation function in the vacuum. To fit our potential to the realistic interaction, we numerically determine $U_S$ and $F_{\rm c}(k)$ to reproduce the $^1S_0$ phase shift $\delta_S(k)$ evaluated by the partial wave analysis from nucleon scattering data, by solving Eq.~(\ref{Eq:pseq}). In this procedure, we do not assume a specific form of $F_{\rm c}(k)$, in contrast to the previous works~\cite{PhysRevA.97.013601,Tajima:2019saw}. Here we mention that as well known the $^1S_0$ interaction becomes repulsive in the high-momentum region, because the sign of $\delta_S(k)$ changes from positive to negative as increasing the scattering energy. In our model, we consider only attractive part by restricting the range of the momentum region, where the $^1S_0$ phase shift is positive. We simply set $F_{\rm c}(k) =0$ where $\delta_S(k) < 0$.
Effects of the repulsive interactions in the high-momentum region are left for our future work.
\par
Figure \ref{Fig:ps} shows the calculated form factor $F_{\rm c} (k)$ with the coupling constant $mU_S/|a_s|=0.79$ as well as the reconstructed $^1S_0$ phase shift. We find that our result completely reproduces $\delta_s(k)$ in the region where $\delta_S(k)> 0$ ($k \le1.78$ fm$^{-1}$ $\equiv k_{\rm{max}}$
). We also compare our results with the ERE method, which is used in the previous work to study the equation of state in the $^1S_0$ neutron superfluidity~\cite{PhysRevA.97.013601}.
 In ERE, the form factor $F_c(k)$ was assumed to be a function with a single cutoff parameter, and the cutoff parameter, as well as $U_S$, is determined to reproduce the $^1S_0$-wave scattering length $a_s=-18.8$ ${\rm fm}$ and the effective range $r_{\rm {eff}}=2.75$ ${\rm fm}$, which characterize the low momentum properties of $\delta_S(k)$. Thus, as shown in Fig.~\ref{Fig:ps} (b), the phase shift estimated within ERE (chain line) gradually deviates from the $^1S_0$ phase shift data as the momentum increases. This disagreement of $\delta_S(k)$ in the high momentum region is improved in our potential. We also mention that in Ref.~\cite{Tajima:2019saw} they overcome this problem by using a multi-rank separable potential including a repulsive part with some cutoff parameters, and investigated the superfluid instability in the normal phase of neutron system above the superfluid transition temperature $T_{\rm c}$.
\par
It has been known that the $^1S_0$ attraction is strong in the low-density region. In the present work, therefore, we take into account superfluid fluctuations within the NSR scheme~\cite{Nozieres1985}, which has been widely used for studying BCS-BEC crossover phenomena in the context of the cold atom physics. For this purpose, it is convenient to employ the path-integral method for the fermionic field $c$ and $\bar{c}$~\cite{PhysRevLett.71.3202}, starting from the partition function
\begin{align}
Z=\int \mathcal{D}
\left[
\bar{c}, c
\right] 
\exp
\left [
-S
\left(
\bar{c}, c
\right)
\right ],
\label{Eq:z}
\end{align} 
with an action $S$ for the Hamiltonian Eq.~\eqref{Eq:Ham} given by
\begin{align}
S\left(
\bar{c}, c
\right)
&=\int_0^{\beta} d\tau
\sum_{{\bm k}\sigma}
\bar{c}_{{\bm k}\sigma}
\left( \tau \right)
\left( \partial_\tau + \xi_k \right)
c_{{\bm k}\sigma}
\left( \tau \right)
\nonumber
\\
&\quad 
+
\sum_{{\bm k},{\bm k}',{\bm q}}
\int_0^{\beta} d\tau
V_{S}\left({\bm k},{\bm k}'\right)
\bar{c}_{{\bm k}+\frac{\bm q}{2}\uparrow}
\left( \tau \right)
\bar{c}_{-{\bm k}+\frac{\bm q}{2}\downarrow}
\left( \tau \right)
c_{-{\bm k}'+\frac{\bm q}{2}\downarrow}
\left( \tau \right)
c_{{\bm k}'+\frac{\bm q}{2}\uparrow}
\left( \tau \right)
\nonumber
\\
&=
S_{\rm{kin}}
\left(
\bar{c}, c
\right)
+S_{\rm{int}}
\left(
\bar{c}, c
\right),
\end{align} 
in the imaginary time formalism with the inverse temperature $\beta=1/T$. As usual, we first introduce a bosonic pairing field $\Phi({\bm q},\tau )$ as an auxiliary field, and apply the Hubbard-Stratonovich transformation for $S_{\rm int}$  as
\begin{align}
e^{-S_{\rm int}}
&= \int {\cal D}[\bar{\Phi},\Phi] 
\exp
\left(
-
\sum_{{\bm q}}\int_0^{\beta} d\tau
\left(
\frac{\beta|\Phi({\bm q},\tau)|^2}{U_S} 
+\sqrt{\beta} \bar{\rho}_S({\bm q},\tau)\Phi({\bm q},\tau) 
+ \sqrt{\beta} \bar{\Phi}({\bm q},\tau) \rho_S({\bm q},\tau) \right)\right),
\end{align}
where $\rho_S({\bm q},\tau)=
\sum_{\bm k}
c_{-{\bm k}+\frac{\bm q}{2}, \downarrow}
\left( \tau \right)
c_{{\bm k}+\frac{\bm q}{2}, \uparrow}
\left( \tau \right)
F_{\rm c}(k)$ and $\sqrt{\beta}$ is multiplied to $\Phi$ and $\bar{\Phi}$ for the normalization. $\bar{\Phi}$ is the complex conjugate of $\Phi$. Integrating out the fermion degrees of freedom in Eq.~(\ref{Eq:z}) and taking the Fourier transformation for $\tau$, we obtain an effective action as 
\begin{align}
S_{\rm eff} &= 
\beta \sum_{{\bm q},i\nu_n} \frac{|\Phi({\bm q},i\nu_n)|^2}{U_S} 
+\sum_{{\bm k},i\omega_l,{\bm k}',i\omega_{l'}} \left(
\beta \xi_{k} \delta\left({\bm k}-{\bm k}' \right)\delta_{l,l'}
- {\rm Tr} \ln \left( 
\beta \hat{G}^{-1}\left({\bm k}~i\omega_l,{\bm k}'~i\omega_{l'}\right) 
\right) \right) ,
\label{Eq:seff}
\end{align}
where $\omega_l=(2l+1)\pi T$ ($\nu_n=2\pi n T$) are the fermionic (bosonic) Matsubara frequency, respectively. $G({\bm k}~i\omega_l,{\bm k}'~i\omega_{l'})$ in Eq.~(\ref{Eq:seff}) is the 2$\times$2 single-particle Green's function defined by
\begin{align}
G^{-1} \left({\bm k}~i\omega_l,{\bm k}'~i\omega_{l'} \right) &=
\left(i\omega_l \sigma_0-\xi_k \sigma_3\right)\delta\left({\bm k}-{\bm k}' \right)\delta_{l,l'}
\nonumber
\\
&+\bar{\Phi} \left({\bm k}-{\bm k}', i\omega_l-i\omega_{l'} \right)
F_{\rm c}\left( \frac{{\bm k}+{\bm k}'}{2} \right)\sigma_- 
+\Phi\left( {\bm k}'-{\bm k}, i\omega_{l'}-i\omega_l \right)
F_{\rm c} \left(\frac{{\bm k}+{\bm k}'}{2} \right)\sigma_+
. 
\end{align}
Here, $\sigma_{\pm}=(\sigma_1\pm\sigma_2)/2$ and $\sigma_{i}$ ($i=0,1,2,3$) are the Pauli matrices acting on the Nambu particle-hole space. The first term describes the kinetic energy (diagonal component of $G^{-1}$) and the second and third terms describe the pairing field (off-diagonal component of $G^{-1}$). The bosonic pairing field $\Phi({\bm q}, i\nu_n)$ is conveniently divided into two parts as 
\begin{equation}
\Phi({\bm q}, i\nu_n) = \Delta\delta({\bm q}) \delta_{i\nu_n,0}+ \delta\Delta({\bm q} ,i\nu_n),
\end{equation}
where $\Delta$ is the saddle point solution and $\delta\Delta({\bm q}, i\nu_n)$ is a fluctuation from $\Delta$.
In the NSR theory, the effective action $S_{\rm {eff}}$ is expanded with respect to $\delta\Delta({\bm q}, i\nu_n)$ up to quadratic order. Then, we obtain
\begin{equation}
S_{\rm{eff}} \simeq S_{\rm{MF}} + \delta S_{\rm{fluct}}.
\label{Eq:snsr}
\end{equation}
Here we express the mean-field contribution as
\begin{align}
S_{\rm MF} &= 
 \frac{ \beta|\Delta|^2}{U_S} 
+ \sum_{{\bm k},i\omega_l} \left(
\beta\xi_k
- {\rm Tr} \ln \left( 
\beta \hat{G}^{-1}_0({\bm k},i\omega_l) 
\right) 
\right), 
\end{align}
with the Green's function within the mean-field theory given by
\begin{equation}
\hat{G}_0^{-1}({\bm k},i\omega_l) = i\omega_l \sigma_0-\xi_k \sigma_3 +
\Delta(k)\sigma_1,
\end{equation}
where we have introduced $\Delta(k) \equiv \Delta F_{\rm c}(k)$ as the momentum dependence in the superfluid order parameter.
We also express the fluctuation contributions as
\begin{equation}
\delta S_{\rm{fluct}}=
\frac{\beta}{2}\sum_{q} 
\Lambda^\dagger ({\bm q}, i\nu_n)
\left(
\frac{1}{U_S} \sigma_0 + \hat{\pi} ({\bm q}, i\nu_n)
\right)
\Lambda ({\bm q}, i\nu_n),
\end{equation}
where $\Lambda^\dagger ({\bm q}, i\nu_n) = (\delta \Delta^\dagger ({\bm q}, i\nu_n), \delta \Delta (-{\bm q}, -i\nu_n))$ is the two-component bosonic field in the Nambu space, and
\begin{align}
\hat{\pi}({\bm q}, i\nu_n)&=
\frac{1}{4}\left(
\begin{array}{cc}
\pi_{11}+\pi_{22}+i\left( \pi_{12}-\pi_{21} \right) & \pi_{11}-\pi_{22} \\
\pi_{11}-\pi_{22} & \pi_{11}+\pi_{22}-i\left( \pi_{12}-\pi_{21} \right)\\
\end{array}
\right)
,
\\
\pi_{ss'}({\bm q}, i\nu_n)&=
\frac{1}{\beta}\sum_p 
\mathrm{Tr} \left(
\sigma_{s} 
\hat{G}_0 \left( {\bm k+\frac{\bm q}{2}},i\omega_l \right)
\sigma_{s'} 
\hat{G}_0\left( {\bm k-\frac{\bm q}{2}}, i\omega_l-i\nu_n \right)
\right)
F_{\rm c}^2 \left( k \right)~~~~\left(s,s'=1,2 \right),
\end{align}
is the 2$\times$2-matrix pair correlation function in the lowest order. We note that $\pi_{11}$ and $\pi_{22}$ denote physically the amplitude and phase fluctuations of the superfluid order parameter, respectively, and  $\pi_{12}$ and $\pi_{21}$ describe the coupling between them. The effective action $S_{\rm eff}$ in Eq.~(\ref{Eq:snsr}) induces the strong-coupling correction to the thermodynamic potential in terms of the thermodynamic relation $\Omega=-T\ln Z=\Omega_{\rm{MF}} + \delta \Omega_{\rm{fluct}}$, where
\begin{align}
\Omega_{\rm{MF}}
&=
\frac{\left| \Delta \right|^2 }{U_S}
+\sum_{\bm k} \xi_k
-\sum_{\bm k} E_k
-2T\sum_{\bm k} \ln\left(1+ e^{-\beta E_{k}} \right),
\\
\delta\Omega_{\rm{fluct}}
&=
\frac{1}{2\beta}
\sum_{q}
\mathrm{Tr} 
\ln \left( 1+U\hat{\pi}({\bm q},i\nu_n) \right).
\end{align}
Here $E_{k}=\sqrt{\xi_k^2+|\Delta(k)|^{2}}$ is the quasiparticle energy spectrum.
\par
In this formalism, the effects of pairing fluctuations are taken into account by self-consistently solving the gap equation together with the particle number equation for $\Delta$ and $\mu$. The gap equation is given by the saddle point condition $(\partial \Omega_{\rm MF}/\partial \Delta)_{N,V}=0$ as
\begin{equation}
\frac{1}{U_S}= \sum_{\bm{k}}
\frac{F_c^2 (k)}{2E_k} 
\tanh\frac{\beta E_k}{2}.
\label{gapeq}
\end{equation}
This equation has the same form as one in the ordinary mean-field theory. The particle number equation is obtained from the thermodynamic relation $N=-(\partial \Omega/\partial \mu)_{V,T}$. When we divide the total particle number $N$ into the mean-field contributions $N_{\rm MF}$ and the strong-coupling correction $\delta N_{\rm fluct}$, we obtain 
\begin{align}
N&=N_{\rm {MF}}+\delta N_{\rm {fluct}},
\label{eqpn}
\\
N_{\rm {MF}}&=\sum_{{\bm k}} \left( 1- \frac{\xi_k}{E_k} \tanh \frac{\beta E_k}{2}\right),
\label{eqnMF}
\\
\delta N_{\rm {fluct}}&=\frac{1}{2\beta} \sum_{{\bm  q},i\nu_n} {\rm{Tr}} \left( 
\hat{\Gamma}\left( {\bm  q},i\nu_n \right) \frac{\partial \hat{\pi} \left({\bm  q},i\nu_n \right) }{\partial \mu}
\right).
\label{eqnfluct}
\end{align}
Here $\hat{\Gamma}$ is the many-body scattering matrix defined by
\begin{align} 
\hat{\Gamma}\left( {\bm  q},i\nu_n \right) = -\frac{U_S}{1+U_S\hat{\pi} \left({\bm  q},i\nu_n \right). }
\end{align}
In Eq.~\eqref{eqnfluct}, we have ignored the term $(\partial \delta\Omega_{\rm fluct}/\partial \Delta)_{T}(\partial \Delta/\partial T)_{V,N}$, which is the higher order correction, for simplicity. Here we also mention that in this formalism, the gap equation Eq.~\eqref{gapeq} does not include the modification of the single-particle spectrum, as well as the screening effects of the interaction due to the finite density, which are important to more quantitatively estimate the superfluuid order parameter $\Delta$ and the superfluid transition temperature $T_{\rm c}$.
\par 
Before closing this section, we mention that $\delta N_{\rm fluct}$ includes the number of the noncondensed bosonic pairs below the superfluid transition temperature $T_{\rm c}$ and that of the preformed Cooper pairs above $T_{\rm c}$, respectively. 
Indeed, $\hat{\Gamma}({\bm q},i\nu_n)$ describes the bosonic collective excitations associated with the phase and amplitude fluctuations of the superfluid order parameter, which are known as the Anderson-Bogoliubov (phase, sound, or phonon) and Higgs (amplitude) modes, respectively. The dispersion relations of these modes are obtained from the pole analysis of the analytically continued $\hat{\Gamma}({\bm q},i\nu_n \to z+i\delta)$, where $z$ is the real energy of these modes and $\delta$ is an infinitely small positive number. In the next section, we will discuss the properties of these collective modes. We note that the gap equation is equivalent to the gapless condition of $\hat{\Gamma}$ (the so-called Thouless criterion ${\rm {det}}\,\hat{\Gamma}^{-1}\left(0,0\right)=0$), that guarantees the existence of the gapless Anderson-Bogoliubov mode in the low-energy region. These topics will be discussed in details in the next section. We mention that $\delta N_{\rm fluct}$ includes not only the superfluid fluctuations but also the modifications of the single-particle spectrum, such as the Hartree-Fock (HF) potential and the mass correction. In Ref.~\cite{PhysRevC.88.054315}, it was pointed out that in the normal phase above $T_{\rm c}$, the pairing fluctuations are overestimated without separately treating the HF potential from $\delta N_{\rm fluct}$ in the high density region. However, in the present study, to avoid numerical difficulties, we simply take into account all the effects of the interaction by $\delta N_{\rm {fluct}}$ given by Eq.~\eqref{eqnfluct} for the whole density region, and the validity of our theoretical framework in the high-density region will be discussed in the next section.

\section{Results}\label{sec:result}
\begin{figure}
\centerline{\includegraphics[width=12cm]{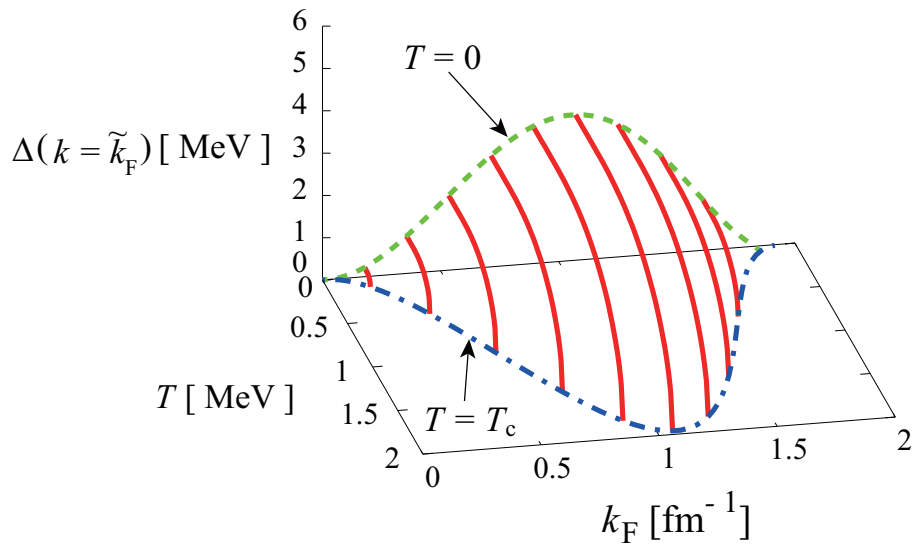}}
\caption{Calculated superfluid order parameter $\Delta(k=\tilde{k}_{\rm F})$ on the effective Fermi surface, where $\tilde{k}_{\rm F}=\sqrt{2m\mu}$, in $^1S_0$ superfluid in neutron stars as functions of the temperature $T$ and the Fermi momentum $k_{\rm F}$. The dashed and the chained line shows the results at $T=0$ and the superfluid transition temperature $T_{\rm c}$, respectively.}
\label{delta3d}
\end{figure}
Figure~\ref{delta3d} shows the superfluid order parameter obtained by self-consistently solving Eqs.~\eqref{gapeq} and \eqref{eqpn} at the effective Fermi surface, which is characterized by the effective Fermi momentum $k=\tilde{k}_{\rm F} \equiv \sqrt{2m \mu}$. The superfluid order parameter is expressed as a function of the temperature $T$ and the Fermi momentum $k_{\rm F}$ (density) in this figure. We first focus on the results at $T=0$. Starting from the low-density region, $\Delta(\tilde{k}_{\rm F})$ gradually increases as the density increases, and has a maximum value around $k_{\rm F} \simeq 1$ ${\rm fm}^{-1}$. Then, $\Delta(\tilde{k}_{\rm F})$ turns to decrease because the interaction strength at the Fermi surface is suppressed due to the form factor $F_{\rm c}(k)$ and finally vanishes at a critical value $k_{\rm F}=1.8$ ${\rm fm}^{-1}$. Note that $F_{\rm c}(k)$ is a decreasing function of $k$ as shown in Fig.~\ref{Fig:ps}. The vanishing of $\Delta$ means that the phase transition from the $^1S_0$ superfluid to the normal state occurs. It was reported that a similar density dependence of the superfluid order parameter was obtained within the mean-field approach~\cite{PhysRevC.57.R1069} and the renormalization group approach~\cite{SCHWENK2003191} with realistic pseudopotentials. 
As shown in Fig.~\ref{delta3d}, a similar density dependence is found in the result for the superfluid transition temperature $T_{\rm c}$, which is consistent with Ref.~\cite{Tajima:2019saw}.
We also briefly note that ERE cannot describe correctly the phase transition, because the attractive potential never vanishes in the high-density region.
\par
\begin{figure}
\centerline{\includegraphics[width=15cm]{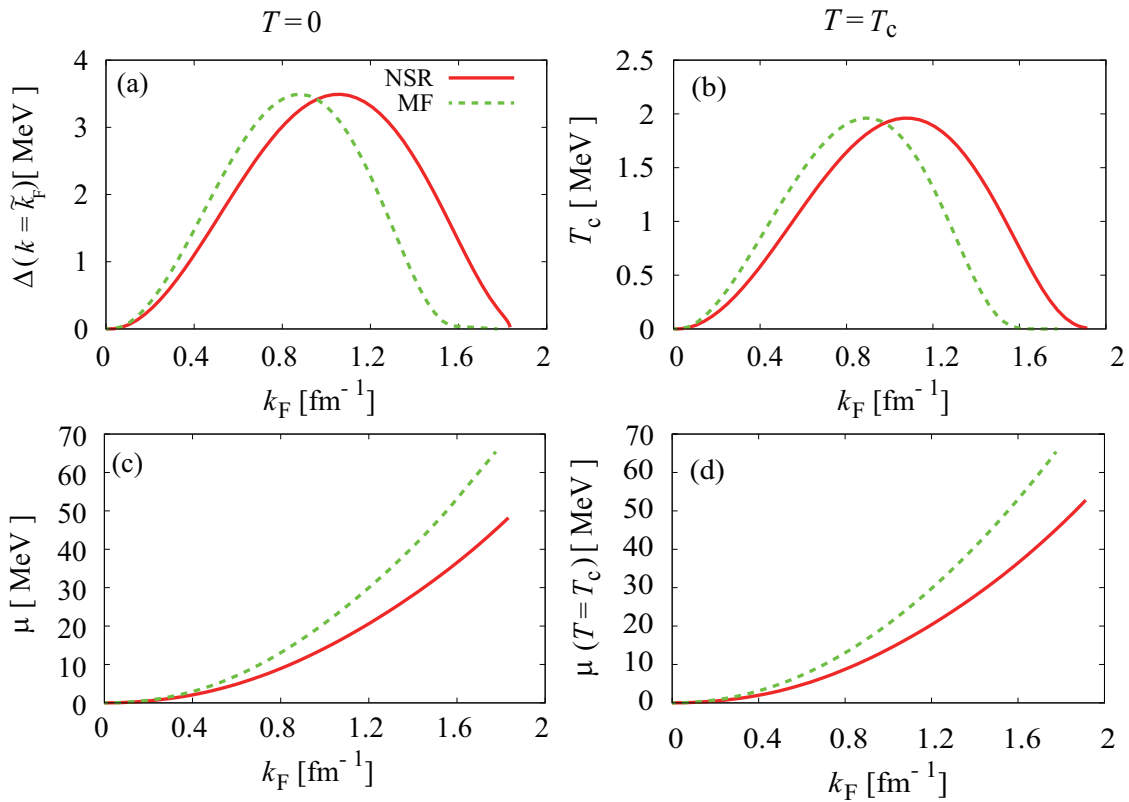}}
\caption{(a) Superfluid order parameter on the effective Fermi surface (where $k=\tilde{k}_{\rm F} \equiv \sqrt{2m\mu}$) at $T=0$, (b) superfluid transition temperature $T_{\rm c}$. In panels, (c) and (d) chemical potential at $T=T_{\rm c}$ and $T=0$ are shown, respectively. In each panel, the results calculated within the mean-field approximation are also shown (dashed lines) for comparison.}
\label{delta}
\end{figure}
In Fig.~\ref{delta}, we compare our results for $\Delta(\tilde{k}_{\rm F})$ at $T=0$ and the superfluid transition temperature $T_{\rm c}$ in the NSR scheme with ones calculated in the mean-field approximation. We mention that in the mean-field theory the particle number equation is obtained by ignoring $\delta N_{\rm fluct}$ in Eq.~\eqref{eqpn}. When we first focus on the low-density region ($k_{\rm F} \lesssim 1~{\mathrm {fm}}^{-1}$), Figs.~\ref{delta}(a) and \ref{delta}(b) show that $\Delta$ and $T_{\rm c}$ are suppressed compared to ones calculated within the mean-field theory due to the superfluid fluctuations. As shown in Figs.~\ref{delta}(c) and \ref{delta}(d), the chemical potential $\mu$ is always suppressed by superfluid fluctuations. This is because in the NSR scheme the noncondensed bosonic pairs are taken into account through the particle number equation, the fermionic component is reduced, and the Fermi sphere is shrunk. Then, the superfluid order parameter also becomes smaller. This behavior is consistent with the ordinary BCS-BEC crossover phenomena in cold atom physics.

\begin{figure}
\centerline{\includegraphics[width=15cm]{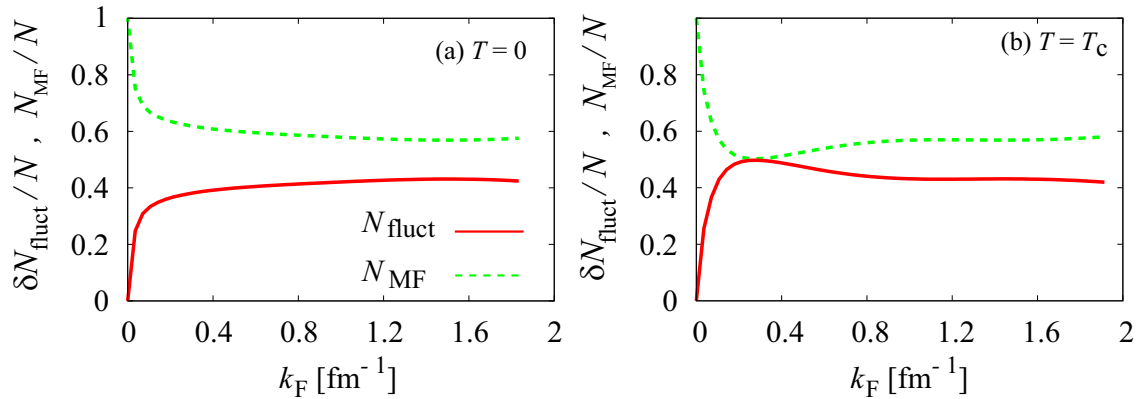}}
\caption{Particle number contribution from the fluctuation term (solid lines) and  the mean-field term (dashed lines) at (a) $T=0$ and (b) $T=T_{\rm c}$.}
\label{pn}
\end{figure}

However, in the higher-density region ($k_{\rm F} \ge\sim 1{\mathrm {fm}}^{-1}$), Figs.~\ref{delta}(a) and \ref{delta}(b) show that the superfluid fluctuations enhance $\Delta$ and $T_{\rm c}$. These enhancements, however, might be an artifact due 
to overestimation of $\delta N_{\rm fluct}$ within our theoretical framework. Figure \ref{pn} shows the calculated density dependence of $N_{\rm MF}$ and $\delta N_{\rm fluct}$ given by Eqs. \eqref{eqnMF} and \eqref{eqnfluct}, respectively, at (a) $T=0$ and (b) $T_{\rm c}$. Starting from the low-density region, $\delta N_{\rm fluct}$ rapidly increases, and eventually approaches some values at the critical density both in the cases at $T=0$ and $T_{\rm c}$. However, as pointed out in Ref.~\cite{PhysRevC.88.054315}, in the normal phase $\delta N_{\rm fluct}$ is improved to be suppressed in the higher-density region by appropriately subtracting the contribution from the HF potential from $\delta N_{\rm fluct}$, and the effects of pairing fluctuations gradually vanish as approaching the critical density where $T_{\rm c} = 0$. As we mentioned in Sec.~ \ref{sec:formalism}, we do not subtract the HF potential to avoid numerical difficulties. The overestimation of $\delta N_{\rm fluct}$ strongly suppresses $\mu$ (see Fig.~\ref{delta}). 
Then, naively one could expect that $T_{\rm c}$, as well as $\Delta$ could also become smaller due to the shrinking Fermi surface. However, because the interaction becomes weaker as increasing the momentum $k$, the suppression of $\mu$ eventually enhances the interaction strength on the effective Fermi surface.
As a result the superfluid order parameter is enhanced in the high-density region.
Although it is still an open question whether in the $^1S_0$ neutron superfluid phase below $T_{\rm c}$ the HF term also qualitatively changes the results, our results in the high-density region might be changed by our theory being improved at this point. We emphasize that, since the HF potential is negligible in the low-density region as discussed in Ref.~\cite{PhysRevC.88.054315}, and $\delta N_{\rm fluct}$ rapidly increases as $k_{\rm F}$ increases, our results indicate that the superfluid fluctuations are important in $^1S_0$ superfluid phase at $T=0$ in neutron stars. 

\begin{figure}
\centerline{\includegraphics[width=18cm]{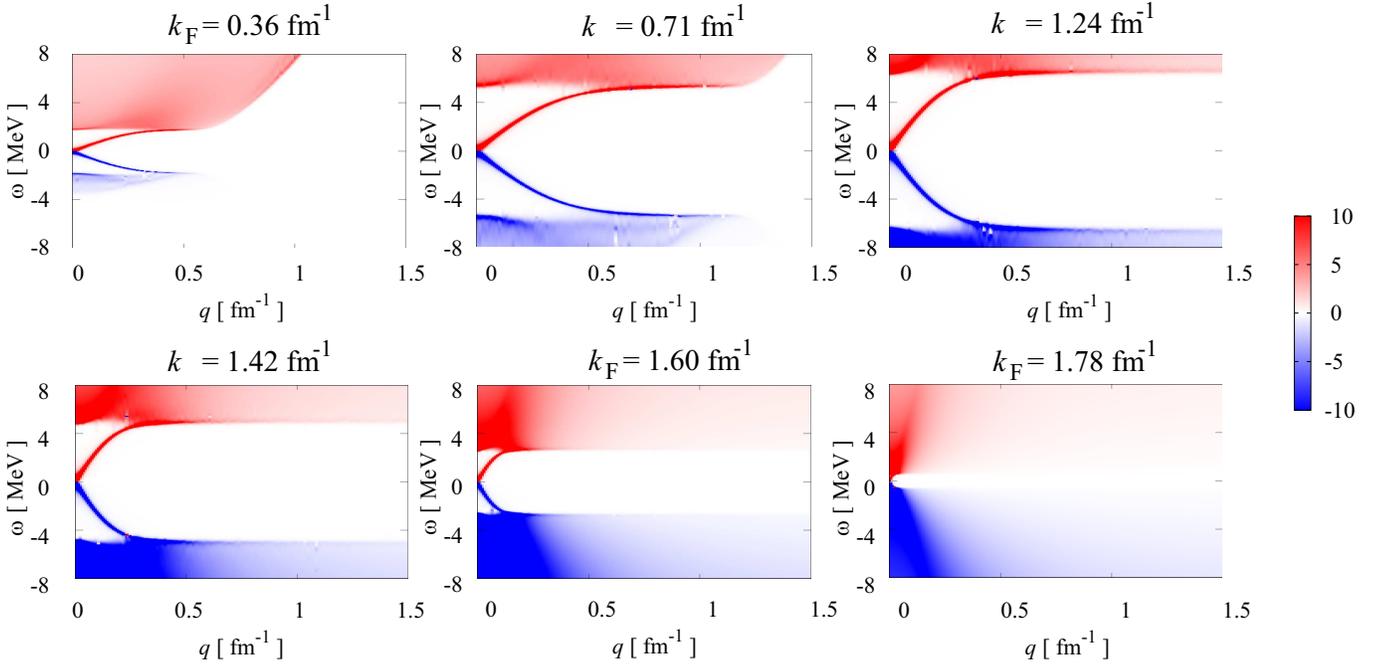}}
\caption{Spectral function $-{\rm Im}\,[\Gamma_{11}({\bm q}, i\nu_n \to \omega+i\delta)]/\pi$ at $T=0$.}
\label{spctgm}
\end{figure}
\begin{figure}
\centerline{\includegraphics[width=9cm]{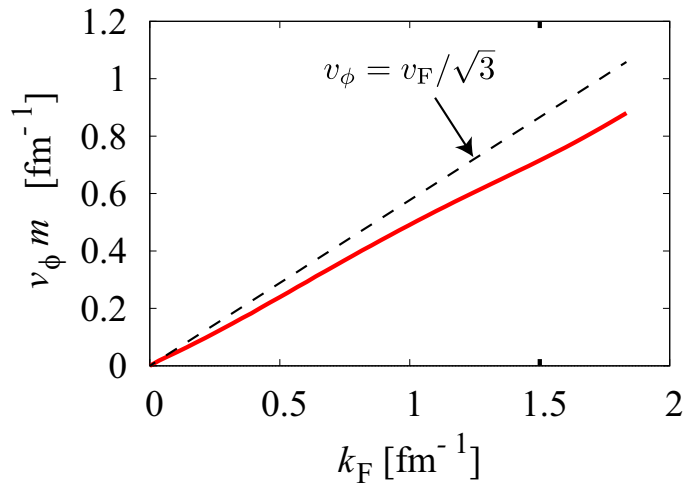}}
\caption{Density dependence of the sound velocity of the Anderson-Bogoliubov mode calculated from Eq.~\eqref{eqvphi} at $T=0$. The dashed line shows the results in the weak-coupling limit $v_{\rm F}/\sqrt{3}$ where $v_{\rm F}$ is the Fermi velocity.}
\label{vphi}
\end{figure}

$\delta N_{\rm fluct}$ affects the thermodynamic properties significantly. To explain this, we first note that, as we mentioned in the previous section, $\delta N_{\rm fluct}$ physically means the number of the noncondensed bosonic pairs, which are dominated by the gapless mode, i.e., the Anderson-Bogoliubov mode having the linear dispersion relations $\omega_q=v_{\phi}q$ with the sound velocity $v_\phi$ in the low temperature limit. Indeed, as shown in Fig.~\ref{spctgm}, we find that the sharp peak structure along $\omega_q=v_{\phi}q$ appears in the spectral function $-{\rm Im}\,[\Gamma_{11}({\bm q}, i\nu_n \to \omega+i\delta)]$ in the whole density region at $T=0$. By expanding the pole condition ${\rm det}\,[\hat{\Gamma}^{-1}({\bm q},i\nu_n \to \omega+i\delta)]=0$ with respect to ${\bm q}$ and $\omega$, the explicit expression of $v_{\phi}$ is obtained as
\begin{align}
v_\phi &=\sqrt{\frac{\eta}{\zeta}},
\label{eqvphi}
\end{align}
with
\begin{align}
\eta&=\sum_{\bm p} 
\left(
\frac{\Delta^2\left( p \right)
}{E_p^3}F_{\rm c}^2\left( p \right)
\right)
\sum_{\bm k} 
\left[
\frac{1}{2E^5_{k}} \left(
\frac{\xi_{k} k}{2m}
+ \frac{ \Delta\left( k \right) \Delta F_c'\left( k \right)}{2}
\right)^2
\right.
\nonumber
\\
&-\frac{1}{2E_k^3}
\left.
\left(
\frac{\xi_{k}}{4m} 
+\left( \frac{k}{2m} \right)^2
+\frac{\Delta^2 F_c\left(k\right)}{12} 
\left(
F_c''\left( k\right)
+
\frac{F_c'\left( 
k \right)
}{k}\right)
+
\left(
\frac{\Delta F_c'\left( k \right) }{2}
\right)^2\right)\right]
F_{\rm c}^2\left( k \right)
,
\\
\zeta&=\sum_{\bm p} 
\left(
\frac{\Delta^2\left( p \right)
}{E_p^3}F_{\rm c}^2\left( p \right)
\right)
\sum_{\bm k} 
\left(
\frac{1}{4E_k^3}
F_{\rm c}^2\left( k \right)
\right)
+
\left(\sum_{\bm k} 
\frac{\xi_k}{2E_k^3}
F_{\rm c}^2\left( k \right)\right)^2
.
\end{align}
Similar results were obtained in the context of cold atom physics~\cite{PhysRevA.67.063612,PhysRevA.74.042717}, as well as nuclear matter~\cite{PhysRevC.90.065805}. Figure~\ref{vphi} shows the sound velocity $v_{\phi}$ of the Anderson-Bogoliubov mode as a function of the Fermi momentum. In the low density limit, $v_{\phi}$ coincides with the expression in weak coupling limit, $v_\phi=v_{\rm F}/\sqrt{3}$. As the density increased, $v_\phi$ gradually deviates from the results in the weak coupling limit, and becomes suppressed due to the strong fluctuations in superfluid pairings. We note that when the momentum dependence of $\Delta(k)$ is ignored, Eq.~\eqref{eqvphi} gives the expression of the sound velocity in the ordinary BCS-BEC crossover~\cite{PhysRevA.67.063612}. 
\par
As pointed out in Ref.~\cite{PhysRevC.90.065805}, the existence of the gapless-collective mode qualitatively changes the thermodynamic properties in the low-temperature limit. Since the single-particle excitations have an energy gap associated with the superfluid order parameter, the contributions from these excitations to thermodynamic quantities are exponentially suppressed as decreasing $T$, as well as developing $\Delta$. However, the gapless Anderson-Bogoliubov mode gives the power low behavior on $T$ to thermodynamic quantities, such as equation-of-state, specific heat, compressibility, and so on~\cite{PhysRevC.90.065805}. Thus, in the low-temperature limit, thermodynamic properties should be dominated by the Anderson-Bogoliubov mode. The collective excitations in $^1S_0$ superfluid in neutron stars were discussed within a theory based on the random phase approximation (RPA) in the previous work~\cite{PhysRevC.90.065805}. However, to quantitatively estimate the effects of the Anderson-Bogoliubov mode in the whole region below $T_{\rm c}$, RPA is not sufficient, because the superfluid fluctuations are not taken into account when one estimates $\Delta$ and $\mu$. Although, we do not calculate the thermodynamic quantities in the present work, the NSR scheme can be applied to the thermodynamics in $^1S_0$ neutron superfluid. It is our future work to investigate the effects of the superfluid fluctuations on the thermodynamic properties.

We comment that the Anderson-Bogoliubov modes are damped in the quasi-particle continuum, which is defined as $\omega^2 \ge {\rm min}_{\bm k} [E_{{\bm k}+{\bm q}/2}+E_{{\bm k}-{\bm q}/2}]^2$, because in this region, the collective excitations decay into two quasiparticle excitations. Reflecting the density dependence of $\Delta(\tilde{k}_{\rm F})$ shown in Fig.~\ref{delta} (a), the structure of the continuum nonmonotonically changes as the density increases. At the critical density of the phase transition from the $^1S_0$-superfluid to the normal state ($k_{\rm F}=1.78~{\rm {fm}}^{-1}$), the Anderson-Bogoliubov mode eventually becomes to be damped in the whole region except at ${\bm q}=0$ and $\omega=0$. We also comment that the amplitude Higgs mode is always located in the quasiparticle continuum and are not clearly seen in the spectral function ${\rm Im}\,\Gamma_{11}$. For this reason, the Higgs mode does not play a crucial role in the thermodynamics near $T=0$. As shown in Fig.~\ref{pn} (a), even in the low-density region ($k_{\rm F} \simeq 0.1~{\rm fm}^{-1}$), $\delta N_{\rm fluct}$ at $T=0$ accounts for about 40\% of the total density of the neutrons. Thus, our results clearly indicate that the superfluid fluctuations should be taken into account for the thermodynamics in $^1S_0$ superfluids in neutron stars.

\section{Summary and discussion} \label{sec:summary}

To summarize, we have discussed the effects of superfluid fluctuations in $^1S_0$ superfluid in neutron stars. To describe the neutron-neutron interaction, we have constructed a separable potential to reproduce the $^1S_0$ phase shift estimated by the partial wave analysis from nucleon scattering data. Using the constructed potential and including superfluid fluctuations within the NSR theory, we have self-consistently determined the superfluid order parameter in a wide density and temperature region. We have found that the superfluid order parameter is suppressed in the low-density region, as a result of the shrunk Fermi sphere due to the suppression of the chemical potential. Although we have found that the superfluid order parameter is enhanced in the high-density region, it might be an artifact of our theoretical framework, because we do not separately treat the HF potential, which is known to be remarkable in the high density region above $T_{\rm c}$, and the contributions from the superfluid fluctuations might be overestimated. This is left as our future work.

We also have shown that the superfluid fluctuations are dominated by the gapless Anderson-Bogoliubov mode with a linear dispersion relation with a sound velocity in the low temperature region. Furthermore, we have found that the contribution from the superfluid fluctuations to the particle number accounts for 40\% of the total number of neutrons even in the low-density region. Since the single-particle excitations are strongly suppressed due to an energy gap associated with the superfluid order parameter, our results indicate that for studying the thermodynamic quantities, such as equation-of-state, specific heat, and compressibility in the neutron stars, the superfluid fluctuations should be taken into account.

The Anderson-Bogoliubov modes studied in this paper are expected to significantly affect the cooling process of neutron stars by neutrino emissions (see Refs.~\cite{Graber:2016imq,Baym:2017whm} and references therein).
Possible impacts of the present study on the cooling process of neutron stars remain as one of important future problems.

In this paper, we have considered only the attractive part of the $^1S_0$ interaction between neutrons. To access the higher-density region, the repulsive part of the $^1S_0$ interaction should be included. In addition, it has been known that the $^3P_2$ attractive interaction also becomes significantly strong as the density increases.  Thus, in more realistic situation, the phase transition from $^1S_0$ to $^3P_2$ superfluid 
or the coexistence of them should be discussed~\cite{Takatsuka1971}. It is in progress to extend our formalism to the case with the $^1S_0$ repulsion as well as the $^3P_2$ attraction.
\par
\par
\acknowledgements
We thank Yoji Ohashi and Hiroyuki Tajima for useful discussions. This work is supported by the Ministry of Education, Culture, Sports, Science (MEXT)-Supported Program for the Strategic Research Foundation at Private Universities ``Topological Science'' (Grant No. S1511006). This work is also supported in part by Japan Society for the Promotion of Science (JSPS) Grant-in-Aid for Scientific Research [KAKENHI Grants No. 17K05435 (S. Y.), No. 16H03984 (M. N.), and No. 18H01217 (M. N.)], and also by MEXT KAKENHI Grant-in-Aid for Scientific Research on Innovative Areas ``Topological Materials Science,'' Grant No. 15H05855 (M. N.).
\par
\bibliography{neutronstar}
\end{document}